# Creating a Discipline-specific Commons for Infectious Disease Epidemiology

Michael M. Wagner, MD, PhD
Dept. of Biomedical Informatics
University of Pittsburgh

Adam Darr
Dept. of Biomedical Informatics
University of Pittsburgh

Alexander T. Loiacono
Department of Health Outcomes and
Biomedical Informatics
University of Florida

William Hogan, MD, MS
Department of Health Outcomes
and Biomedical Informatics
University of Florida

Matt Diller
Department of Health Outcomes
and Biomedical Informatics
University of Florida

Terence Sperringer, Jr.
Dept. of Biomedical Informatics
University of Pittsburgh

John Levander
Dept. of Biomedical Informatics
University of Pittsburgh

Max Sibilla
Dept. of Biomedical Informatics
University of Pittsburgh

Shawn T. Brown, PhD
Hewlett-Packard Enterprise

Abstract:

**Objective:** To create a commons for infectious disease (ID) epidemiology in which epidemiologists, public health officers, data producers, and software developers can not only share data and software, but receive assistance in improving their interoperability.

**Materials and Methods:** We represented 586 datasets, 54 software, and 24 data formats in OWL 2 and then used logical queries to infer potentially interoperable combinations of software and datasets, as well as statistics about the FAIRness of the collection. We represented the objects in DATS 2.2 and a software metadata schema of our own design. We used these representations as the basis for the Content, Search, "FAIR-o-meter," and Workflow pages that constitute the MIDAS Digital Commons.

**Results:** Interoperability was limited by lack of standardization of input and output formats of software. When formats existed, they were human-readable specifications (22/24; 92%); only 3 formats (13%) had machine-readable specifications. Nevertheless, logical search of a triple store based on named data formats was able to identify scores of potentially interoperable combinations of software and datasets.

**Discussion:** We improved the findability and availability of a sample of software and datasets and developed metrics for assessing interoperability. The barriers to interoperability included poor documentation of software input/output formats and little attention to standardization of most types of data in this field.

**Conclusion:** Centralizing and formalizing the representation of digital objects within a commons promotes FAIRness, enables its measurement over time and the identification of potentially interoperable combinations of data and software.

## 1. OBJECTIVE

This paper describes how we created the MIDAS Digital Commons (MDC), which is a discipline-specific commons for infectious disease (ID) epidemiology. The scope of MDC is defined by the scope of research conducted by the Models of Infectious Disease Agent Studies (MIDAS) research network [1]. This one network represents most disciplines ranging from biomedical sciences to ecology involved in research in ID epidemiology, with a common focus on using computational models of disease transmission.

The MDC consists of (1) software, datasets, and data formats that we have made as compliant as possible with FAIR (Findable, Accessible, Interoperable, and Re-usable) guiding principles; (2) an ontology-based indexing and search capability for these objects; and (3) a cluster compute platform named *Olympus*. Our inclusion of software and data formats as digital objects in MDC alongside datasets was an outgrowth of our strong emphasis on interoperability in a field where *in silico* experiments are de rigueur.

Our objective was to create a commons in which epidemiologists, public health officers, data producers, and software developers could not only share datasets and software, but also improve their interoperability. Due to the need for rapid response to pandemic diseases, adoption of data-science innovations can be expected to benefit both applied and basic science.

## 2. BACKGROUND AND SIGNIFICANCE

Despite the big-data characteristics of ID epidemiology, the field has not given the same attention as have the 'omics' to making its datasets and software as reusable as possible. Figure 1 illustrates the problem in more detail, beginning with primary datasets being produced about populations, their environments, diseases, and pathogens by governmental and non-governmental organizations around the world. The information is encoded (when encoded at all) in diverse formats, necessitating substantial workflows of data extraction, cleansing, integration, and interpretation. With the exception of information about sequences and pathogens,[1] the findability and accessibility of these datasets are suboptimal and interoperability and re-use are low. We discuss Figure 1 in more detail in subsequent sections.

A particular challenge for ID epidemiology is marshalling the datasets for an epidemic in a particular location quickly enough to obtain analytic results that make a difference. There are many recent examples of this challenge for epidemics of H1N1 Influenza [2 3], Ebola [4 5], Chikungunya [6-8], Dengue [9], and Zika [10]. The requirement is not only that needed data exist, but also that they are findable, accessible… in short … FAIR.

The 2017 winner of the Open Science Prize— *Real-Time Evolutionary Tracking for Pathogen Surveillance and Epidemiological Investigation*—is an excellent example of the potential of open science in ID epidemiology [11]. The winning project develops the nextflu and nextstrain systems, which are both available at nextstrain.org. These systems would not be possible without the open sharing of sequence data. And their maximal impact depends on open sharing in near real time, especially during epidemics. A workflow of the nextstrain system is depicted in Figure 1, bottom.

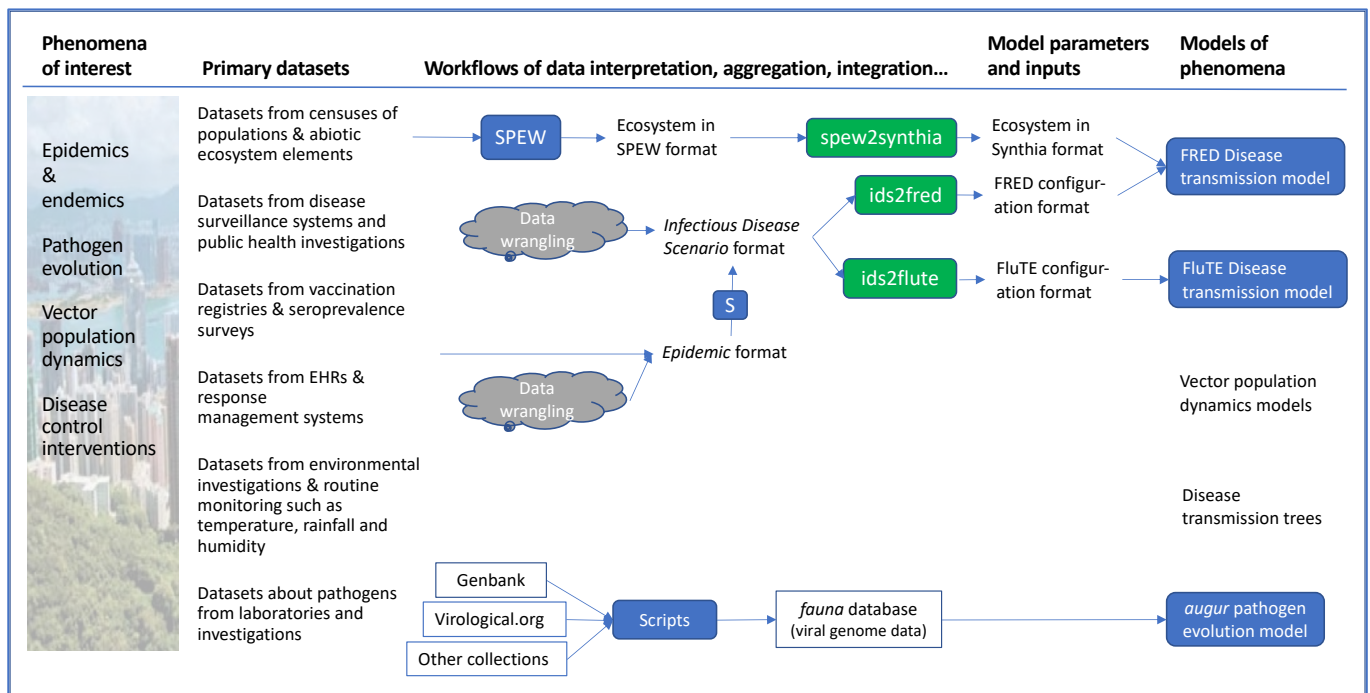

Figure 1. The data scientist's view of ID epidemiology, using examples from MIDAS. The biological phenomena of interest to infectious disease epidemiologists—epidemics, endemic diseases, pathogen evolution, disease control interventions—take place in populations within ecosystems. Thus, the primary datasets in this domain include not only data about pathogens, infections and infectious diseases, but also censuses of populations and ecosystem elements such as forests, skyscrapers, rivers, the atmosphere and food and water supplies that are involved in disease transmission and control. *Legend*: clear rectangle, data store; rounded rectangle, software; green, data-format converter; *ids2fred*, converts format from Apollo-encoded infectious disease scenario to FRED input; *SPEW*, Synthetic Populations and Ecosystems of the World; *S,* unnamed software.

---

[1] The Bioinformatics Resource Centers funded by NIAID are web-based resources anchored in sequence data. Their websites contain datasets, tools, search functions and in some cases workflow support. As such, they can be thought of a proto-commons for applied research on organisms considered potential agents of biowarfare or bioterrorism or causing emerging or re-emerging diseases. Of particular relevance to ID epidemiology are: the Virus Pathogen Resource (ViPR) [12 13], Influenza Research Database [14 15], Pathosystems Resource Integration Center (PATRIC) for bacteria [16-18], Eukaryotic Pathogen Database Resources [19] and VectorBase [20].

## 2.1 FAIR Commons

As part of its Big Data to Knowledge (BD2K) initiative, the US National Institutes of Health (NIH) is supporting pilot projects of a resource-sharing device known as a *commons* [21]. [22] defines a *commons* as "[an] interoperable infrastructure that co-locates data, storage, and computing infrastructure with common analysis tools." An NIH commons additionally adds requirements that data, software and other objects of research be Findable, Accessible, Interoperable, and Reusable (FAIR) [23]; that the compute platform consist of cloud and/or high-performance computing resources; that datasets and software be uniquely and persistently identified; and that there be user interfaces to the commons (see for example, Figure 1 in [24] and Slide 14 in [25]). An experienced operator of commons has identified additional requirements to maximize the reuse of digital objects by machines [22].

Metadata play a key role in making objects FAIR. BD2K's bioCADDIE project developed the DATS v2.2 metadata schema [26] for use in its prototype *DataMed* indexing and retrieval system [27]. In the MDC described in this paper, we use the DATS v2.2 *Dataset* and *DataStandard* types for metadata about datasets and data formats. BD2K's CEDAR project is creating tools to minimize the effort required to annotate digital objects with metadata, which we expect to adopt in our work [28].

The subject of metadata for software has received attention in a BD2K workshop [29], a recent review by the Force11 Software Citation group identifying multiple software metadata schemas developed by numerous communities [30], and the CodeMeta Project [31], which has created a mapping between these software metadata schemes and an exchange format that potentially allows software repositories to interoperate. However, we elected to develop our own metadata types for software as current approaches are not expressive enough for input and output formats, which are key determinants of interoperability.

## 2.2 Improving access and ease of use of disease transmission models (DTMs)

Although there are many important models in ID epidemiology, disease transmission models (DTMs) stand out for covering the key biological process of interest: pathogen transmission. Therefore, in prior work, we developed the workflow shown in Figure 1 that goes from an instance of *Infectious Disease Scenario*—a complex data type—through one or more data format converters into the native inputs of one or more DTMs [32-34]. To partly address the effort of data wrangling to create an instance of *Infectious Disease Scenario*, we defined an additional complex data type *Epidemic*, which is a more proximate target for data wrangling from primary data. One of our key goals for the MDC is to further wring data wrangling from that workflow and others.

## 2.3 Ontology-based representation of digital objects

Our ontological analysis of the science and practice of infectious disease epidemiology began in 2011 with a focus on epidemics and in silico models of epidemics. It then extended to populations and ecosystems, and presently the scope has extended to the nature of datasets and software depicted in Figure 1. We have expressed the results of this analysis in Apollo-SV, an OWL 2 ontology [32], which we have used as the basis for an ontology-based catalog of software, datasets, publications, and grey literature in ID epidemiology.[2] We note that other than indexing a relatively small number of datasets and software objects thus making them findable via a SPARQL query of the triple-store database, our approach made no specific efforts to make the software and datasets more widely discoverable, interoperable, and re-usable which is the focus of the MDC reported here.

The MDC uses Apollo-SV just like the Neurosciences Information Framework (NIF) uses NIFSTD, its suite of ontologies, as the basis for indexing and retrieval of datasets and software [35]. Like NIFSTD, Apollo-SV re-uses pre-existing ontologies as much as possible, and brings all ontologies together under the Basic Formal Ontology as the upper layer. MDC like NIF then leverages the OWL 2 logic-based representation to enhance search for digital objects.

## 3. MATERIALS AND METHODS

Our design of the commons was guided by ontological analysis, FAIR Guidelines [36], suggested requirements for a data commons in [22], and the following use cases.

---

[2] The OBC.ide query interface, its dynamically-generated disease information pages, and the report of all publications are available at http://ide.onbc.io/#/.

### 3.1 Use cases

1. As a public-health official, I want to find all DTMs and disease forecasters that can be used in my jurisdiction for inclusion in a report.
2. As an infectious-disease epidemiologist, I want to find and access all software, datasets, publications and reports about epidemic disease D in location L in order to build or adapt a DTM or disease forecasting model for D in L.
3. As a data scientist, I want to find combinations of datasets and software that match on their input and output data formats to identify potential workflows.
4. As a data scientist, I want statistics and diagnostics about the objects in my commons that are not FAIR so that I can improve their reusability and potential for interoperability.
5. As an epidemiologist, public health officer, govt. agency or software developer, I would like to not only share datasets and tools, but also ensure that those digital objects can be used by other entities, both human and machine.
6. As another commons, data discovery index, or software discovery index, I want to discover the set of digital objects in MDC and to access their metadata and data, to either increase their discoverability or reuse the objects.

### 3.2 Sources of digital objects

We elicited software and datasets from MIDAS network researchers and included DTMs and Apollo-encoded datasets from our previous work. We added data formats (as objects) that were used by the included software and datasets.

### 3.3 Digital object identifiers (DOI)

We obtained DOIs for our own objects and encouraged other creators to obtain DOIs or permit us to obtain them on their behalf from Zenodo [37]. In the absence of a DOI, we formed a unique identifier as the combination of a namespace that the creator or we control and a unique identifier within that namespace. For datasets still under development, such as the Zika epidemic datasets, we have temporarily overloaded the identifier field with 'identifier will be created at time of release.'

For software, we recognize that there are use cases for assigning identifiers anywhere along the spectrum from product line (e.g., Microsoft Word) to a software release, to an installed instance [38], but at present we only uniquely identify source code releases.

### 3.4 Metadata representations of datasets, data formats and software

Our process for creating metadata is shown in Figure 2. We represented datasets and data formats per the DATS 2.2 *Dataset* and *DataStandard* schemas, respectively [39]. When representing data formats, we used the *DataStandard* schema's *extraProperties* field to add "human-readable specification of data format," "machine-readable specification of data format," and "validator." The values of these properties are IRIs to the documentation, schema or validator software.

We generated dataset and data format metadata semi-automatically. For example, a Python program created metadata for each of 363 Project Tycho datasets from a listing of only the property values that changed from dataset to dataset (e.g., identifier, country, condition, dates, and access URL). The same program generated metadata for 122 SPEW synthetic ecosystems and 88 Apollo-encoded epidemics. The program automated the assignment of identifiers and alternative identifiers for locations by calling the Apollo Location Service API, which is a data service in the MDC.

We represented software per the MDC Software Metadata XML Schema Document [40], which includes the 12 minimal metadata fields promulgated in Bonnazi, 2015 [29]. We represented software input and output data formats as lists of type string, where the strings were either a data format identifier or N/A (not available). We added elements to the base Software type as necessary for the 11 subtypes of *Software*. For example, *Disease transmission model* and *Disease forecasting algorithm* add elements for locations and pathogens for which the software is designed.

Due to the uniqueness of software, we generated software metadata manually.

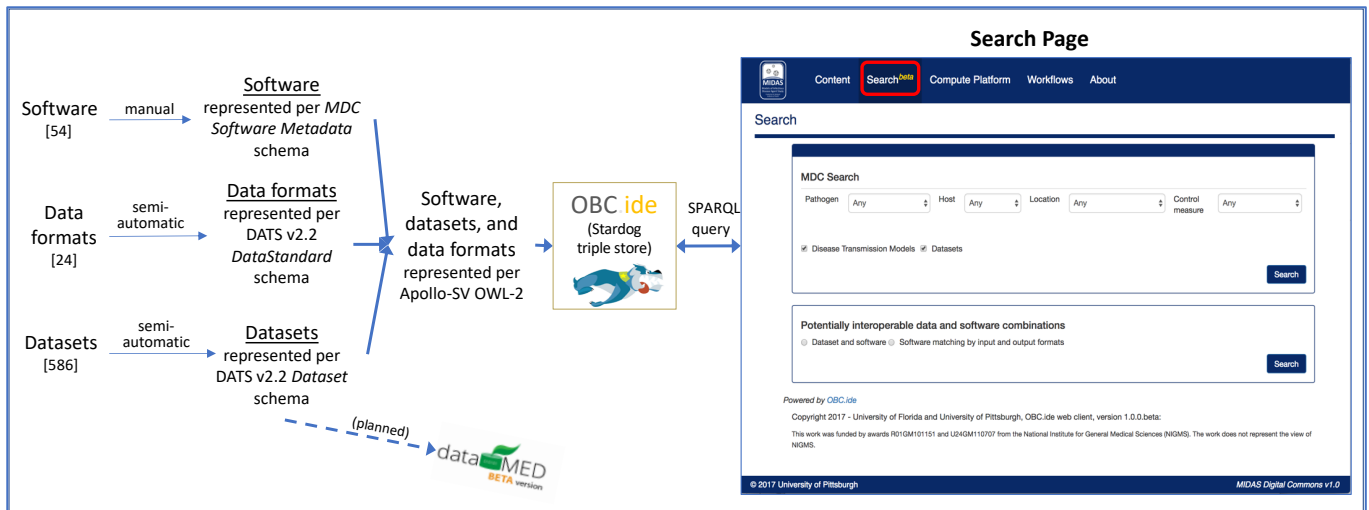

Figure 2. Making objects findable. We first represent datasets and data formats using DATS schema v 2.2 and software using our own XML Schema Definition for software. We then represent the objects in OWL 2 prior to storage in the OBC.ide triple store, which can be searched by SPARQL queries. For datasets, the DataMed dataset discovery index will be a second path by which they become findable.

### 3.5 OWL 2 representations of digital objects

We also created OWL 2 representations of each digital object, which then became the basis for a triple store that can be searched by SPARQL queries (Figure 2).

This process involved developing ontological representations of digital objects and then writing software that automated most of the conversion from JSON (either DATS or our software XML metadata converted to JSON) to OWL 2 individuals. The remaining, manual parts of the process involved mapping XML and JSON attribute values to ontology classes (for things like control measures) and creating additional OWL 2 instances required for indexing the digital objects. The final output of the overall semi-automated process was a set of OWL 2 files that we loaded in bulk into the triple store used by OBC.ide.

### 3.6 Search

We designed the search capabilities of MDC to satisfy use cases 1-4; specifically, search for DTM by host, pathogen, location, control measures; find all digital objects plus publications and reports about a specific epidemic; interoperability queries; counting numbers and computing percentages of digital objects meeting certain fair principles (can do a query counting how many have a DOI, another query for how many software objects have a non-proprietary input format, ditto output format, plus analogous counts of datasets, for example).

### 3.7 Compute Platform

The MDC uses a dedicated high-performance cluster called *Olympus* located at the Pittsburgh Supercomputing Center. Olympus hosts seven DTMs and 122 synthetic ecosystems and is configured with several programming languages, compilers, and popular development tools for general modeling work, as described in the MDC's *Compute Platform* tab. For each digital object that is available on the Olympus Cluster, the Content Page displays an *AOC* indicator.

## 4. RESULTS

### 4.1 Content

Figure 3 shows the Content Page, which displays all 664 software, datasets, and data formats in the MDC—54 software, 586 datasets, and 24 data formats. Clicking an object brings up a human-readable representation of its metadata designed to give a user what he or she likely needs to determine if the object is relevant, with the option to toggle to the metadata encoding (Figure 4).

Figure 3. *Content Page* of the MDC. The Content page displays the entire set of digital objects in browsable expand-collapse trees. Those objects available on the Olympus Cluster are flagged *AOC;* DTMs available via a universal user interface are flagged *UIDS*; datasets encoded in machine-interpretable Apollo XSD types are flagged *AE;* and objects requiring sign-on for access are flagged *SSO*. The page also displays a list of standard identifier systems to encourage their use.

Figure 4. Metadata pop-up screens. When a human user of the MDC clicks on an object on the content page, he or she sees the HTML view and the option to switch to a Metadata view. The elements *Pathogen coverage, Location coverage, Host species included,* and *Control measures* are specific to DTMs.

The software objects fall into 11 subtypes (Table 1) that vary not only in purpose, but in the kinds of information they require as input, the kind of information they output (if any), and the kinds of software that they may typically interoperate with.

The most common type of software in the MDC—*DTMs*—require the widest range of input information. Software in the *Data visualizer* category is often the last link in a chain of processing, so typically has no output format. *Modeling platforms* are software that are agnostic to type of model, so have no infectious disease informational inputs or outputs.

The set of software includes the R packages HyPhy, spaero and Seedy, which contain multiple programs that may fall into multiple software types. We treat R packages as single objects at present and do not attempt to characterize their input and output data formats in the metadata, although we may do so in the future. Similarly, we treat the *Disease forecasters* Dengue forecasting project and Flusight—which involve multiple teams' forecasts—as single objects and do not attempt to characterize the different data inputs used by different competitors.

Table 1. Types of Software

| Type [number of instances] | Information inputs | Information outputs | Types of software they may follow |
|---|---|---|---|
| Disease transmission models [18] | Population immunity and infection levels, infection transmission parameters, treatment efficacies, disease control measures | Number infected, treated, dead over time and space, attack rates, effective $R_0$ | Data visualizers |
| Disease forecasters [8] | Disease surveillance data, unconventional. | Forecasts of future incidence in time and space, peak dates of epidemics | Data visualizers |
| Pathogen evolution models [4] | Pathogen gene sequences | Phylogenetic trees | Data visualizers |
| Population dynamics models [2] | Initial spatial and age distributions, life cycle, factors that influence mortality and morbidity | Future spatial and age distributions, mortality and morbidity statistics | Data visualizers, Disease transmission models, disease forecasters |
| Synthetic ecosystem constructors [1] | Census products | Individuals, households, places (e.g., schools) | Disease transmission models |
| Disease transmission tree estimators [1] | Results of case investigations +/- sequences isolated from cases | Transmission tree | Data visualizers |
| Phylogenetic tree constructors [2] | Pathogen sequences | Phylogenetic trees | Data visualizers |
| Data visualizers [6] | Diverse kinds of information | — | Any software with output |
| Data-format converters [2] | Diverse kinds of information | Same information as input | Any software with output |
| Data services [6] | Diverse kinds of information | Similar to input, but may be integrated or standardized | — |
| Modeling platforms [4] | — | — | — |

The dataset objects do not fall into subtypes, per se, since DATS does not recognize subtypes. However, the vast majority of datasets can be naturally grouped by their data formats into (1) 52 synthetic ecosystems encoded in the SPEW.US format, (2) 69 synthetic ecosystems encoded in the SPEW.IPUMS format, (3) 363 disease surveillance datasets encoded in the Tycho Dataset Format, and (4) 88 epidemics encoded as instances of the Apollo XSD *Epidemic* type. There are a small number of "datasets" without formats; they represent health department websites that provide access to multiple datasets that we have yet to formalize.

## 4.2   Identifiers

Of 664 objects, 10 (1.5%) had preexisting DOIs. Of the remaining 654, we used the Zenodo API to obtain identifiers for 122 synthetic ecosystems, 27 Ebola epidemics, 3 rabies case series, and the H1N1 infectious disease scenario, leaving 501 objects without DOIs. We delayed obtaining DOIs for 424 datasets because the eternal nature of the DOI and associated dataset made their developer want to recheck them.

## 4.3 Adherence to FAIR principles

The FAIR-o-meter shown in Figure 5 is a dynamically created report of statistics about the FAIRness of digital objects in the MDC. We plan to use these numbers, and breakdowns by specific subtypes of software and data, to guide improvements in interoperability.

The SPARQL queries that generate the FAIR-o-meter statistics are publicly available in a GitHub repository with the search SPARQL queries discussed below.

In general, software and data formats are currently less FAIR than datasets. The majority of software objects do not have DOIs or documented input and output formats. The six data visualizers are an exception: all had at least one defined input format and averaged 2 per visualizer.

Of 24 data formats, 22 (92%) had human-readable specifications, but only 3 (13%) had machine-readable (XSD) specifications and an additional two had validators, meaning that for 18 of 24 (75%) data formats there is no way for a machine to check conformance.

| Attribute | Software | Datasets | Total |
| --- | --- | --- | --- |
| Digital Object Identifier | 10 (19%) | 153 (26%) | 163 (25%) |
| Any identifier | 11 (20%) | 525 (90%) | 536 (84%) |
| Named data format | 23 (43%) | 579 (99%) | 602 (94%) |
| Software with at least one input format | 19 (35%) | Not applicable | Not applicable |
| Software with at least one output format | 11 (20%) | Not applicable | Not applicable |
| Software with both input and output format | 7 (13%) | Not applicable | Not applicable |
| Reusable license | 20 (37%) | 577 (98%) | 597 (93%) |
| Named data format and reusable license | 12 (22%) | 577 (98%) | 589 (92%) |
| Total | 54 (100%) | 586 (100%) | 640 (100%) |

Figure 5. FAIR-o-meter

## 4.4 OWL 2 representations of digital objects

The OWL 2 representations of digital objects include OWL individuals for not just the digital objects themselves, but also for their associated entities (e.g., source code repository, DOI, website, and license). The reason was to take advantage of the logical reasoning inherent in OWL for querying and to avoid ad-hoc OWL 2 annotations and annotation values (e.g., strings and literals) that frustrate interoperability due to lexical variation, synonymy, ambiguity/homonymy, and typographical errors. Even identifiers—especially DOIs—are OWL individuals, which enables us easily to ask the question: *does this digital object have an identifier, and if so what kind of identifier(s) and what are their properties?* A simplistic annotation approach would have required understanding which OWL annotation properties represent identifiers and which do not, knowledge that would then have been inaccessible to machines (outside the semantics of RDF/OWL).

At present, the search queries leverage the taxonomic and part-of reasoning capabilities of OWL 2. For example, searching for models that simulate influenza returns models indexed as simulating influenza A. Searching for models that simulate disease transmission in North America retrieves models that simulate in any subpart of North America.

## 4.5 Search

We developed SPARQL queries for (1) generating the content page, (2) searching for digital objects based on geographical location, pathogen, host, and infectious disease control measure, and (3) using data format information to query for sets of software and data that are potentially interoperable (Table 2). The second query retrieves both datasets and disease transmission models potentially relevant to a modeling study (Figure 6 shows a query for Zika in South America). The goals of the third query are to identify potential workflows and to encourage software and dataset developers to implement data standards. The large number of dataset–software–software triples (218) stems from the large number of datasets with a named data format relative to software.

Figure 6. Search screen

Table 2. (potential) Digital object interoperability identified by OBC queries[*]

| Digital object interoperability patterns | N | Examples |
|---|---|---|
| Software A–Software B pair | 16 | Ebola transmission model, Galapagos v1.0 data visualizer |
| Dataset–Software–Software triple | 218 | Florida Synthetic Ecosystem, spew2synthia data format converter, pFRED disease transmission model |

[*]Assumptions: (1) software have a single input and a single output format, (2) the format is the only constraint that needs to be met for the digital objects to be interoperable, and (3) data and software are fully conformant to the format.

## 4.6 Workflows

We define workflows that execute on Olympus with scripts, as illustrated by the *Obtaining a Synthetic Population in Synthia Format* workflow found on the Workflow Page. This workflow is also shown in Figure 1. We also define workflows that execute on the Apollo Web Services such as a workflow that begins with an instance of *Epidemic* and ends with a visualization of the output of a DTM.

## 4.7 API

The MDC API is a RESTful API that (1) returns a list of the DOIs in MDC, (2) returns an object's metadata when presented with a DOI, and (3) returns data when presented with a DOI and a distribution identifier (we specify one our more distributions of a dataset in our metadata) of a dataset. We documented the MDC API using the Swagger software library [41], which describes the endpoint URLs, HTTP methods, required parameters, optional parameters, response syntaxes and input forms to test the API directly from a web browser.

## 5. DISCUSSION

We built a FAIR commons for infectious disease epidemiology, a discipline with a challengingly diverse range of data and an equally challenging number of governmental and non-governmental entities around the world involved in their collection.

We populated the commons with an initial set of datasets and software, which we plan to expand rapidly. We focused on increasing their interoperability by making explicit their data formats, developing metrics of FAIRness and queries that find potentially interoperable chains of datasets and software within the MDC. Our very simple representation of software input and outputs allows us to identify potential interoperability and is also showing what other information is needed for more precise identification of interoperable combinations of datasets and software.

There are public health data repositories (e.g., Project Tycho, the Centers for Disease Control WONDER, or the World Health Organization Global Health Observatory) and increased open sourcing of software using github and R (CRAN), but few if any infectious disease epidemiology commons that aspire to be an NIH style commons with an explicit commitment to FAIR compliance. This lack of commitment could seriously limit opportunities for public health to benefit from data-driven analytics for discovery that are emerging in molecular and clinical biomedical sciences such as cancer genomics [42]. To construct and sustain public health repositories that are FAIR compliant will require investments in human and technical resources. In return, a commons can arguably make more efficient use of data and software resources for the common good. A commons can make clear what standards are in use in which countries or agencies and allow competing standards to exert their gravitational effects for users. A commons can also motivate data creators by showing how their data is a basis for beneficial (to science and the public's health) analytic use.

In the next phase, we plan to focus on rapid expansion of the software and datasets and to increase interoperation with data repositories and resources cited in the background section. We feel expansion is the most important factor in success and have added staff for it. We also plan to work on two related basic problems that haven't received sufficient attention: formalizing workflows as objects so machines can reason about them and representing the data inputs and outputs of software for the same reason.

If the FAIR Commons approach to increasing sharing, standardization, and interoperability succeeds in this domain, as we expect it will, we can expect to see successful applications similar to that of nextstrain for forecasting and disease transmission modeling.

## 6. CONCLUSION

We demonstrated how a scientific field can create a discipline-specific commons by first identifying its key biological phenomena and models of interest, followed by assembling a set of core software and datasets. Our use cases, design ideas, and software may be useful to those considering creating a commons for other disciplines.

We also demonstrated ways to measure the relative interoperability of the objects in the collection as a means to improve interoperability by exposing formats, creating data-format conversion software, and ultimately standardization.

## 7. ACKNOWLEDGMENTS

Wilbert van Panhuis, MD, PhD, Jessi Espino, MD, MS, Amanda Hicks, PhD, Jay DePasse,
Tonatiuh Mendoza and Sahawut Wesaratchakit contributed to this work.

This work was funded by awards U24GM110707 and R01GM101151 from the National Institute for General Medical Sciences (NIGMS), and award UL1TR001427 from the National Center for Advancing Translational Sciences (NCATS). This paper does not represent the view of NIGMS or NCATS.